\documentstyle[prl,aps,epsfig,floats]{revtex}

\newcommand\beq{\begin{equation}}
\newcommand\eeq{\end{equation}}
\newcommand\bea{\begin{eqnarray}}
\newcommand\eea{\end{eqnarray}}
\newcommand\nonum{\nonumber}
\newcommand\bi{\bibitem}

\begin{document}

\draft

\textheight=23.8cm
\twocolumn[\hsize\textwidth\columnwidth\hsize\csname@twocolumnfalse\endcsname

\title{\Large Gapless Phase in the $XYZ$ Spin-1/2 Chain in a Magnetic 
Field and the Quantum ANNNI Chain}
\author{Amit Dutta$^1$ and Diptiman Sen$^2$} 
\address{\it $^1$ Max-Planck-Institut f\"ur Physik komplexer Systeme, 
N\"othnitzer Strasse 38, 01187 Dresden, Germany \\
$^2$ Centre for Theoretical Studies,
Indian Institute of Science, Bangalore 560012, India}

\date{\today}
\maketitle

\begin{abstract}
We study the $XYZ$ spin-1/2 chain placed in a magnetic field pointing along 
the $x$-axis. We use bosonization and a renormalization group analysis to 
show that the model has a non-trivial fixed point at a certain value of 
the $XY$ anisotropy $a$ and the magnetic field $h$. Hence, there is a line 
of critical points in the plane $(a,h)$ on which the system is gapless,
even though the Hamiltonian has no continuous symmetry. 
The quantum critical line separates a gapped commensurate phase from a gapped 
incommensurate phase. Our study explains why the floating phase of the 
axial next-nearest neighbor Ising (ANNNI) chain in a transverse magnetic field
is only a line, as shown by recent numerical studies.
\end{abstract}
\vskip .5 true cm

\vskip.5pc
]
\vskip .5 true cm

One-dimensional quantum spin systems have been studied extensively ever since
the isotropic spin-1/2 chain was solved exactly by Bethe. Baxter
later used the Bethe ansatz to solve the fully anisotropic ($XYZ$) spin-1/2 
chain in the {\it absence} of a magnetic field \cite{baxter}; the problem
has not been analytically solved in the presence of a magnetic field.
Experimentally, quantum spin chains and ladders are known to exhibit a wide 
range of unusual properties, including both gapless phases with a power-law 
decay of the two-spin correlations and gapped phases with an exponential decay
\cite{gogolin,dagotto}. There are also two-dimensional classical statistical 
mechanics systems (such as the axial next-nearest neighbor Ising (ANNNI) 
model) whose finite temperature properties can be understood by studying an
equivalent quantum spin-1/2 chain 
in a magnetic field. The ANNNI model has been studied by several techniques,
and it was believed for a long time to have a floating phase of finite width
in which the system is gapless \cite{selke}. However, a recent large-scale 
numerical study has shown that the width of this phase seems to be zero 
within numerical errors \cite{shirahata}.

Amongst the powerful analytical methods now available for studying 
quantum spin-1/2 chains is the technique of bosonization \cite{gogolin}.
Recently, the $XXZ$ chain in a magnetic field \cite{dmitriev} and the quantum 
ANNNI model \cite{allen} have been studied using bosonization. In this work, 
we will study the fully anisotropic $XYZ$ model in a magnetic field pointing 
along the $x$-axis. For small values of the $XY$ anisotropy $a$ and the 
magnetic field $h$, we will show that there is a non-trivial fixed point 
(FP) of the renormalization group (RG) in the $(a,h)$ plane; the system is 
gapless on a quantum critical line of points which flow to this FP. 
The floating phase of the ANNNI model will be shown to be a 
special case of our results corresponding to $\Delta =0$. The gapless line is 
somewhat unusual because the $XY$ anisotropy and the magnetic field both
break the continuous symmetry of rotations in the $x-y$ plane. We will 
provide a physical understanding of this line by going to the classical 
(large $S$) limit of the model.

We consider the Hamiltonian defined on a chain of sites
\bea
H ~=~ \sum_n ~[~ & & (1+a) ~S_n^x S_{n+1}^x ~+~ (1-a)~S_n^y S_{n+1}^y \nonum \\
& & +~ \Delta S_n^z S_{n+1}^z ~-~ h S_n^x ~]~.
\label{ham1}
\eea
We will assume that the $XY$ anisotropy $a$ and the $zz$ coupling $\Delta$
satisfy $-1 \le a, \Delta \le 1$. We can assume without loss of generality 
that the magnetic field strength $h \ge 0$. For $a=h=0$, the model is 
symmetric under rotations in the $x-y$ plane and is gapless. The low-energy 
and long-wavelength modes of the system are described by the bosonic 
Hamiltonian \cite{gogolin}
\beq
H_0 ~=~ \frac{v}{2} ~\int ~dx ~[~ (\partial_x \theta)^2 ~+~ (\partial_x 
\phi)^2 ~]~,
\label{ham2}
\eeq
where $v$ is the velocity of the low-energy excitations (which have the
dispersion $\omega = v |k|$); $v$ is a function of $\Delta$. 
(The continuous space variable $x$ and the site label $n$ are related through 
$x=nd$, where $d$ is the lattice spacing). The bosonic theory contains 
another parameter called $K$ which is related to $\Delta$ by \cite{gogolin}
\beq
K ~=~ \frac{\pi}{\pi + 2 \sin^{-1} (\Delta)} ~.
\label{kd}
\eeq
$K$ takes the values $1$ and $1/2$ for $\Delta =0$ (which describes 
noninteracting spinless fermions) and $1$ (the isotropic antiferromagnet) 
respectively; as $\Delta \rightarrow -1$, $K \rightarrow \infty$. We thus 
have $1/2 \le K < \infty$. 

The spin-density operators can be written as \cite{dmitriev}
\bea
S_n^z ~&=&~ {\sqrt {\frac{\pi}{K}}} ~\partial_x \phi ~+~ (-1)^n c_1
\cos (2 {\sqrt {\pi K}} \phi ) ~, \nonum \\
S_n^x ~&=&~ [~ c_2 \cos (2 {\sqrt {\pi K}} \phi) ~+~ (-1)^n c_3 ~]~ 
\cos ({\sqrt {\frac{\pi}{K}}} \theta ) ~, 
\label{spin}
\eea
where the $c_i$ are constants given in Ref. 10. The $XY$ anisotropy term is
given by
\beq
S_n^x S_{n+1}^x ~-~ S_n^y S_{n+1}^y ~=~ c_4 \cos (2{\sqrt {\frac{\pi}{K}}} 
\theta ) ~,
\label{anis}
\eeq
where $c_4$ is another constant. 

For convenience, let us define the three operators
\bea
{\cal O}_1 ~&=&~ \cos (2 {\sqrt {\pi K}} \phi) \cos ({\sqrt {\frac{\pi}{K}}}
\theta ) ~, \nonum \\
{\cal O}_2 ~&=&~ \cos (2 {\sqrt {\frac{\pi}{K}}} \theta ) ~, \quad {\rm and}
\quad {\cal O}_3 ~=~ \cos (4 {\sqrt {\pi K}} \phi) ~.
\label{ops}
\eea
Their scaling dimensions are given by $K + 1/4K$, $1/K$ and $4K$ respectively.
Using Eqs. (\ref{spin}-\ref{anis}), the terms corresponding to $a$ and $h$ in 
Eq. (\ref{ham1}) can be written as
\beq
H_a + H_h ~=~ \int dx ~[~ a c_4 {\cal O}_2 - h c_2 {\cal O}_1 ~]~ ,
\eeq
where we have dropped rapidly varying terms proportional to $(-1)^n$ since 
they will average to zero in the continuum limit. (We will henceforth absorb 
the factors $c_4$ ($c_2$) in the definitions of $a$ ($h$)). 
We will now study how the parameters $a$ and $h$ flow under RG.

The operators in Eqs. (\ref{ops}) are related to each other through the 
operator product expansion; the RG equations for their coefficients 
will therefore be coupled to each other \cite{cardy}. In our model, 
this can be derived as follows. Given
two operators $A_1 = \exp (i\alpha_1 \phi + i\beta_1 \theta)$ and $A_2 =
\exp (i\alpha_2 \phi + i\beta_2 \theta)$, we write the fields $\phi$ and 
$\theta$ as the sum of slow fields (with wave numbers $|k| < \Lambda e^{-dl}$) 
and fast fields (with wave numbers $\Lambda e^{-dl} < |k| < \Lambda$), where 
$\Lambda$ is the momentum cut-off of the theory, and $dl$ is the change in 
the logarithm of the length scale. Integrating out the fast fields shows that
the product of $A_1$ and $A_2$ at the same space-time point gives a
third operator $A_3 = e^{i(\alpha_1 + \alpha_2 )\phi + i(\beta_1 + \beta_2 )
\theta}$ with a prefactor which can be schematically written as
\beq
A_1 A_2 ~\sim ~ e^{-(\alpha_1 \alpha_2 + \beta_1 \beta_2 ) dl /2\pi} ~A_3 ~.
\eeq
If $\lambda_i (l)$ denote the coeffients of the operators $A_i$ in an
effective Hamiltonian, then the RG expression for $d\lambda_3 /dl$ will contain
the term $(\alpha_1 \alpha_2 + \beta_1 \beta_2 ) \lambda_1 \lambda_2 /2\pi$.
Using this, we find that if the three operators in (\ref{ops}) have 
coefficients $h$, $a$ and $b$ respectively, then the RG equations are
\bea
\frac{dh}{dl} ~&=&~ (2 - K - \frac{1}{4K}) h ~-~ \frac{1}{K} ah ~-~ 4Kbh ~,
\nonum \\
\frac{da}{dl} ~&=&~ (2 -\frac{1}{K}) a ~-~ (2K -\frac{1}{2K}) h^2 ~, \nonum \\
\frac{db}{dl} ~&=&~ (2 - 4K) b ~+~ (2K -\frac{1}{2K}) h^2 ~, \nonum \\
\frac{dK}{dl} ~&=&~ \frac{a^2}{4} ~-~ K^2 b^2 ~,
\label{rg}
\eea
where we have absorbed some factors involving $v$ in the variables $a$, $b$ 
and $h$. (We will ignore the RG equation for $v$ here). It will turn out that 
$K$ renormalizes very little in the regime of RG flows that we will be 
concerned with. Eqs. (\ref{rg}) have appeared earlier in the context of some 
other problems \cite{nersesyan,giamarchi}. However, the last two 
terms in the expression for $dh/dl$ were not presented in Ref. 8; these two 
terms turn out to be crucial for what follows. Note that Eqs. (\ref{rg}) are 
invariant under the duality transformation $K \leftrightarrow 1/4K$ and 
$a \leftrightarrow b$.

Let us now consider the fixed points of Eqs. (\ref{rg}). For any value of 
$K=K^*$, a trivial FP is $(a^* ,b^* ,h^* )=(0,0,0)$. Remarkably, it turns out
that there is a non-trivial FP for any value of $K^*$ lying in 
the range $1/2 < K^* < 1 + {\sqrt 3}/2$; we will henceforth restrict our 
attention to this range of values. (The upper bound on $K^*$ comes from the
condition $2-K^*-1/4K^* > 0$). The non-trivial FP is given by
\bea
h^* ~&=&~ \frac{\sqrt {2K^*(2-K^*-1/4K^*)}}{(2K^*+1)} ~, \nonum \\
a^* ~&=&~ (K^*+\frac{1}{2}) h^{*2} ~, \quad {\rm and} \quad b^* ~=~ 
\frac{a^*}{2K^*} ~.
\label{fp}
\eea
The system is gapless at this FP as well as at all points which flow to this 
FP. One might object that Eqs. (\ref{rg}) can only be trusted if $a$, $b$ 
and $h$ are not too large, otherwise one should go to higher orders. 
We note that the FP approaches the origin as $K^* \rightarrow 1 
+ {\sqrt 3}/2 \simeq 1.866$; this corresponds to the $zz$ coupling $\Delta 
= - \sin [\pi ({\sqrt 3} - 3/2)] \simeq - 0.666$. Thus the RG equations 
can certainly be trusted for $K^*$ close to $1.866$. For $K^*=1$, the FP is 
at $(a^* , b^* , h^*) = (1/4,1/8, 1/{\sqrt 6})$.

We have numerically studied the RG flows given by Eqs. (\ref{rg}) for various
starting values of $(K,a,b,h)$. Since the Hamiltonian in 
(\ref{ham1}) does not contain the operator ${\cal O}_3$, we set $b=0$ 
initially. We take $a$ and $h$ to be very small initially, and see which
set of values flows to a non-trivial FP. For instance, starting with 
$K=1$, $b=0$, and $a,h$ very small, we find that there is a line of points 
which flow to a FP at $(K^* , a^*, b^*, h^*) = (1.020, 0.246, 0.122, 
0.404)$. This line in the $(a,h)$ plane is shown in Fig. 1. We see 
that $K$ changes very little during this flow; if we start with a larger value
of $K$ initially, then it changes even less as we go to the non-trivial
FP. It is therefore not a bad approximation to ignore the flow of
$K$ completely.

%\vspace*{-0.2cm}
\begin{figure}[htb]
\begin{center}
\hspace*{-0.5cm}
\epsfig{figure=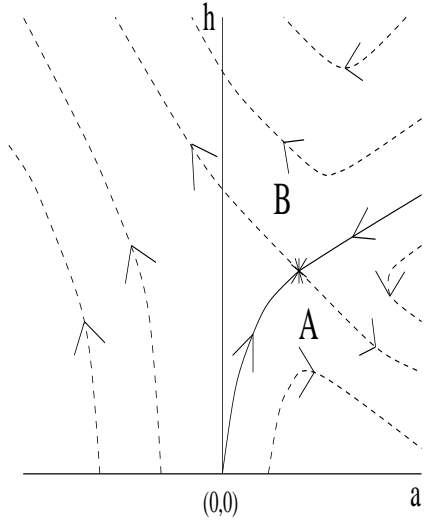,width=7.5cm,height=6cm,angle=270}
\end{center}
%\vspace*{-0.2cm}
\caption{RG flow diagram in the $(a,h)$ plane. The solid line shows the set of
points which flow to the FP at $a^*=0.246$, $h^*=0.404$ marked by an asterisk.
The dotted lines show the RG flows in the gapped phases A and B (see text).}
\end{figure}

We can characterize the set of points $(a,h)$ lying close to the 
origin which flow to the non-trivial FP. Numerically, we find that there is 
a unique flow line in the $(a,h)$ plane for each starting value of $K$ and 
$b=0$, provided that $a, h$ are very small initially. This means 
that $a(l)$ and $h(l)$ given by Eqs. (\ref{rg}) must follow the same line 
regardless of the starting values of $a,h$. From Eqs. (\ref{rg}), we see that 
if $h << a^{1/2}$, then $h(l) \sim h(0) \exp (2-K-1/4K)l$ while $a(l) \sim 
\exp (2-1/K)l$. Hence $h$ must initially scale with $a$ as
\beq
h ~\sim ~a^{(2-K-1/4K)/(2-1/K)} ~,
\label{scaling}
\eeq
as we have numerically verified for $K=1$. 
However, Eq. (\ref{scaling}) is only true if $(2-K-1/4K)/(2-1/K) > 1/2$, i.e.,
if $K < (1 + {\sqrt 2})/2 =1.207$. For $K \ge 1.207$ (i.e., $\Delta \le
-0.266$), the initial scaling form is given by $h \sim a^{1/2}$. 

We now examine the stability of small perturbations 
away from the fixed points. The trivial FP
at the origin has two unstable directions ($a$ and $h$),
one stable direction ($b$) and one marginal direction ($K$). The non-trivial
FP has two stable directions, one unstable direction and a marginal
direction (which corresponds to changing $K^*$ and simultaneously $a^*$, $b^*$
and $h^*$ to maintain the relations in Eqs. (\ref{fp})). The presence of two 
stable directions implies that there is a two-dimensional surface of 
points (in the space of parameters $(a,b,h)$) which flows to this FP;
the system is gapless on that surface. A perturbation in the unstable 
direction produces a gap in the spectrum. For 
instance, at the FP $(K^* ,a^* ,b^* ,h^*)=(1,1/4,1/8, 1/{\sqrt 6})$,
the four RG eigenvalues are given by $1.273$ (unstable), $0$ (marginal), and
$-1.152 \pm 1.067 i$ (both stable). The positive eigenvalue corresponds to
an unstable direction given by $(\delta K, \delta a, \delta b, \delta h)= 
\delta a (0.113, 1, -0.092, -0.239)$. A small perturbation of size $\delta a$ 
in that direction will produce a gap in the spectrum which scales as 
$\Delta E \sim |\delta a|^{1/1.273} = |\delta a|^{0.786}$; the correlation 
length is then given by $\xi \sim v/\Delta E \sim |\delta a|^{-0.786}$.

Fig. 1 shows that the set of points which do not flow to the non-trivial
FP belong to either region A or region B. These regions can be 
reached from the non-trivial FP by moving in the unstable direction, with 
$\delta a > 0$ for region A, and $\delta a < 0$ for region B. In region A, 
the points flow to $a=\infty$; this corresponds to a gapped phase which is 
commensurate since the magnetic field $h$ asymptotically flows to zero. 
In region B, both $a$ and $h$ flow to
$-\infty$; this is a gapped incommensurate phase. The line
separating the two regions is a commensurate-incommensurate transition line.
An order parameter which distinguishes between regions A and B is the
staggered magnetization in the $y$ direction, defined as
\beq
m_y ~=~ [~ \lim _{n \rightarrow \infty} ~(-1)^n < S_0^y S_n^y > ~]^{1/2} ~.
\label{order}
\eeq
This is zero in the commensurate phase A. In the incommensurate phase B, it 
is non-zero, and its scaling with the perturbation $\delta a$ can be found 
as follows \cite{dmitriev}. At $a=h=0$, the leading term in the long-distance 
equal-time correlation function of $S^y$ is given by
\beq
< S_0^y ~S_n^y > ~\sim ~ \frac{(-1)^n}{|n|^{1/2K}} ~.
\eeq
Hence the scaling dimension of $S_n^y$ is $1/4K$. In a gapped
incommensurate phase in which the correlation length is much larger than
the lattice spacing, $m_y$ will therefore scale with the gap as $m_y \sim 
(\Delta E)^{1/4K}$. If we assume that the scaling dimension of $S_n^y$ at the
non-trivial FP remains close to $1/4K$, then the numerical
result quoted in the previous paragraph for $K=1$ implies that $m_y \sim 
|\delta a|^{0.196}$ for $\delta a$ small and negative.

We will now apply our results to the one-dimensional spin-1/2 quantum ANNNI 
model \cite{selke,allen}, with nearest neighbor ferromagnetic and next-nearest
neighbor antiferromagnetic Ising interactions and a transverse magnetic field.
The Hamiltonian is given by
\beq
H_A ~=~ \sum_i ~[~ - 2 J_1 S_i^x S_{i+1}^x + J_2 S_i^x S_{i+2}^x + 
\frac{\Gamma}{2} S_i^y ~]~,
\label{annni1}
\eeq
where $J_1 , J_2 >0$; we can assume without loss of generality that
$\Gamma \ge 0$. The quantum Hamiltonian in (\ref{annni1}) is related
to the transfer matrix of the two-dimensional classical ANNNI model; the
finite temperature critical points of the latter are related to the 
ground state quantum critical points of (\ref{annni1}), with the temperature 
$T$ being related to the magnetic field $\Gamma$. 
Earlier studies showed that the model has a floating phase of finite width 
which is gapless and incommensurate \cite{selke}. A recent 
bosonization study reached the same conclusions 
\cite{allen}. However, a more recent numerical study involving very large 
two-dimensional classical systems at finite temperature concluded that the
floating phase seems to have zero width \cite{shirahata}. 
We can now understand this as follows.
Consider a Hamiltonian which is dual to (\ref{annni1}) for spin-1/2;
this will turn out to be a special case of our earlier model. The dual 
Hamiltonian is given by \cite{selke,sen}
\beq
H_D ~=~\sum_n ~[~J_2 T_n^x T_{n+1}^x ~+~\Gamma T_n^y T_{n+1}^y -J_1 T_n^x ~]~,
\label{annni2}
\eeq
where $T_n^{\alpha}$ are the spin-1/2 operators dual to $S_i^{\alpha}$. Let 
us scale this Hamiltonian by a factor to make $J_2 + \Gamma =2$. 
Then this has the same form as in Eq. (\ref{ham1}), with $J_2 = 1+a$, $\Gamma 
= 1-a$, $J_1 =h$, and $\Delta =0$. Hence it follows that the quantum
ANNNI model has a line of points in the $(J_2 - \Gamma , J_1)$ plane 
on which the system is gapless. (From Eq. (\ref{scaling}), we see that
the shape of this line is given by $J_1 \sim (J_2 - \Gamma)^{3/4}$ as $J_1 
\rightarrow 0$). This line can be identified with the floating phase.
Our results for the width of the floating agree completely with those obtained
for the two-dimensional classical ANNNI model in Ref. 5. We should point out, 
however, that Ref. 5 shows the phase transition to be of the 
Kosterlitz-Thouless type (with $\xi$ diverging exponentially) from the 
high-temperature side (i.e., from region B in Fig. 1), and is of the 
commensurate-incommensurate type (with $\xi$ diverging as a power-law) 
from the low-temperature side (region A in Fig. 1). Our
analysis indicates that it should be of the commensurate-incommensurate type
from both sides. This discrepancy remains to be resolved. One comment to be
made here is that Ref. 5 has studied two values of $J_2 /J_1$ ($0.6$ and
$0.8$) which are not large; our RG results may be expected to be valid 
if $a,h$ are small, i.e., if $J_2 /J_1$ is large.

Finally, we would like to provide a physical picture of the gapless line in 
the $(a,h)$ plane by looking at the classical limit of Eq. (\ref{ham1}). 
Consider the Hamiltonian 
\bea
H_S ~=~ \sum_n ~[~ & & (1+a) ~S_n^x S_{n+1}^x ~+~ (1-a)~S_n^y S_{n+1}^y 
\nonum \\
& & +~ \Delta S_n^z S_{n+1}^z ~-~ h S S_n^x ~]~,
\label{hams}
\eea
where the spins satisfy ${\bf S}_n^2 = S(S+1)$, and we are interested in the
classical limit $S \rightarrow \infty$ \cite{sen}. Let us assume that
the $zz$ coupling is smaller in magnitude than the $xx$ and $yy$ couplings.
Then the classical ground state of (\ref{hams})
is given by a configuration in which all the spins lie in the $x-y$ plane,
with the spins on odd and even numbered sites pointing respectively at an 
angle of $\alpha_1$ and $-\alpha_2$ to the $x$-axis.
The ground state energy per site is 
\bea 
e (\alpha_1 , \alpha_2 ) ~=~ S^2 ~[ & & - \frac{h}{2} (\cos \alpha_1 + 
\cos \alpha_2 ) 
\nonum \\
& & + \cos (\alpha_1 + \alpha_2 ) + a \cos (\alpha_1 - \alpha_2 ) ].
\eea
Minimizing this with respect to $\alpha_1$ and $\alpha_2$, we discover that 
there is a special line given by $h^2 ~=~ 16 a$ on which {\it all} solutions 
of the equation 
\beq
h \cos (\frac{\alpha_1 + \alpha_2}{2} ) ~=~ 4a \cos (\frac{\alpha_1 - 
\alpha_2}{2} ) ~,
\label{gnst}
\eeq
give the same ground state energy per site, $e_0 =-(1+a)S^2$. The solutions of
Eq. (\ref{gnst}) range from $\alpha_1 = \alpha_2 = \cos^{-1} (4a/h)$ to
$\alpha_1 = \pi, \alpha_2 = 0$ (or vice versa); in the ground state phase 
diagram of the ANNNI model, these two configurations correspond to the 
paramagnetic incommensurate phase and the antiphase respectively 
\cite{selke}. Classically, we see that the line $h^2 = 16a$ separates 
these two phases; we therefore identify this line with the floating phase
\cite{sen}. The fact that there is a one-parameter set of ground states 
(characterized by, say, the value of $\alpha_1$ in the solution of 
(\ref{gnst})) which are all degenerate for $h^2 = 16a$ means that there is
a gapless mode in the excitations; we can derive this explicitly by going 
to the next order in a 
$1/S$ expansion. Hence the model is gapless on that line. This provides some 
understanding of why one may expect such a gapless line in the spin-1/2 model 
also. Note however that the bosonization analysis gives the scaling form in 
(\ref{scaling}) for $h$ versus $a$; this agrees with the classical form only 
if $\Delta \le -0.266$.

To summarize, we have shown that the $XYZ$ spin-1/2 chain in a magnetic field 
exhibits a gapless phase on a particular line. It may be interesting 
to use numerical techniques like the density-matrix renormalization group 
method \cite{pat} to study various ground state properties of this model, in 
particular the behavior of order parameters such as the one defined in Eq. 
(\ref{order}). The RG equations studied in this paper appear in other strongly
correlated systems, such as the problem of two spinless Tomonaga-Luttinger 
chains with both one- and two-particle interchain hoppings \cite{nersesyan},
and one-dimensional conductors with spin-anisotropic electron interactions 
\cite{giamarchi}. The gapless phase may therefore also appear in other systems.

D.S. would like to thank P. Fulde for hospitality in the Max-Planck-Institut 
f\"ur Physik komplexer Systeme, Dresden, during the course of this work.
We thank R. Narayanan for useful comments.

\end{document}